\documentclass[conference]{IEEEtran}
\IEEEoverridecommandlockouts

\usepackage{booktabs}
\usepackage{graphicx}
\usepackage{booktabs}
\usepackage{pifont}
\usepackage{makecell}
\usepackage[table]{xcolor}

\usepackage{booktabs}
\usepackage{enumitem}
\usepackage{array}
\usepackage{wrapfig}
\usepackage{amsmath,amsfonts}
\usepackage[noend]{algpseudocode}
\usepackage{graphicx}
\usepackage{textcomp}
\usepackage{float}
\usepackage{listings}
\usepackage{xspace}
\usepackage{multirow}
\usepackage{amsthm}

\usepackage{balance}
\usepackage{algorithm}
\usepackage{algpseudocode}
\usepackage{colortbl}

\usepackage[skins]{tcolorbox}
\usepackage{xcolor,pifont}
\usepackage{multicol}
\newcommand*\colourcheck[1]{%
	\expandafter\newcommand\csname #1check\endcsname{\textcolor{#1}{\ding{52}}}%
}
\colourcheck{blue}
\colourcheck{green}
\colourcheck{red}
\newtcolorbox{boxB}[2][]{%
  enhanced,colback=white,colframe=black,coltitle=black,
  sharp corners,
  toprule=1.0pt,
  rightrule=0.3pt,
  leftrule=0pt,
  bottomrule=0pt,
  fonttitle=\itshape\scshape\large,
  left=0pt,right=5pt,top=5pt,bottom=3pt,
  attach boxed title to top right={yshift=-0.3\baselineskip-0.4pt,xshift=-5mm},
  boxed title style={tile,size=minimal,left=0.2mm,right=0.5mm,
    colback=white,before upper=\strut},
  title=#2,#1
}

\newcommand{\tool}{\textsc{PAIR-Bench}\xspace}

\def\BibTeX{{\rm B\kern-.05em{\sc i\kern-.025em b}\kern-.08em
    T\kern-.1667em\lower.7ex\hbox{E}\kern-.125emX}}

\newboolean{showcomments}
\setboolean{showcomments}{true}
\ifthenelse{\boolean{showcomments}}
 { \newcommand{\mynote}[2]{
      \fbox{\bfseries\sffamily\scriptsize#1}
        {\small$\blacktriangleright$\textsf{\emph{#2}}$\blacktriangleleft$}}}
        { \newcommand{\mynote}[2]{}}

\usepackage{tikz}

\newcolumntype{L}[1]{>{\raggedright\arraybackslash}p{#1}}

\usepackage{amsthm}
\definecolor{dkgreen}{rgb}{0,0.6,0}
\definecolor{gray}{rgb}{0.5,0.5,0.5}
\definecolor{lightgray}{rgb}{211, 211, 211}
\definecolor{mauve}{rgb}{0.58,0,0.82}

\definecolor{custom-blue}{rgb}{0,0,0}
\definecolor{custom-red}{rgb}{1,0,0}
\definecolor{my-blue}{rgb}{0,0,1}

\definecolor{c1}{HTML}{f4cccc}
\definecolor{c2}{HTML}{f5cdcd}
\definecolor{c3}{HTML}{fffcfc}
\definecolor{c4}{HTML}{ffffff}
\definecolor{c5}{HTML}{ffffff}
\definecolor{c6}{HTML}{fffdfd}
\definecolor{c7}{HTML}{f5cfcf}
\definecolor{c8}{HTML}{fffbfb}
\definecolor{c9}{HTML}{ffffff}
\definecolor{c10}{HTML}{fffdfd}
\definecolor{c11}{HTML}{fefafa}
\definecolor{c12}{HTML}{fef7f7}
\definecolor{c13}{HTML}{ffffff}
\definecolor{c14}{HTML}{fffefe}
\definecolor{c15}{HTML}{ffffff}
\definecolor{c16}{HTML}{fefafa}
\definecolor{c17}{HTML}{fdf3f3}
\definecolor{c18}{HTML}{fffefe}
\definecolor{c19}{HTML}{fdf5f5}
\definecolor{c20}{HTML}{ffffff}

\makeatletter
\newcommand{\linebreakand}{%
  \end{@IEEEauthorhalign}
  \hfill\mbox{}\par
  \mbox{}\hfill\begin{@IEEEauthorhalign}
}
\makeatother

\begin{document}

\title{Benchmarking Code Improvement with Progressive, Adaptive, and Interactive Feedback}


\author{\IEEEauthorblockN{Cuong Chi Le}
\IEEEauthorblockA{\textit{University of Texas at Dallas} \\
Texas, USA \\
cuong.le@utdallas.edu}
\and
\IEEEauthorblockN{Aashish Yadavally}
\IEEEauthorblockA{\textit{University of Central Florida} \\
Florida, USA \\
aashish.yadavally@ucf.edu}
\and
\IEEEauthorblockN{Minh Le-Anh}
\IEEEauthorblockA{\textit{FPT Software AI Center} \\
Vietnam \\
minhla4@fpt.com}
\and
\IEEEauthorblockN{Tien N. Nguyen}
\IEEEauthorblockA{\textit{University of Texas at Dallas} \\
Texas, USA \\
tien.n.nguyen@utdallas.edu}
}

\maketitle

\begin{abstract}
Large language models (LLMs) are typically evaluated on code generation and program repair using binary functional correctness: a generated program or patch either passes or fails a test suite. This protocol is simple but coarse, as it ignores partial progress, feedback use, regressions, and the refinement trajectory through which models often improve code. We introduce \tool, a progressive and adaptive benchmark for evaluating \emph{code improvement}: transforming an incorrect or incomplete program into a more correct one through feedback-guided refinement. \tool uses progressive hinting, a structured feedback protocol with two controls. \emph{Failure-region control} determines what the feedback targets by grouping hidden failing tests into failure scenarios, while \emph{hint-depth control} determines how much repair-relevant information is revealed, from coarse symptoms to implementation-level guidance. This design enables \tool to measure whether a model repairs targeted failures, generalizes beyond the hint, preserves already-correct behavior, and how much assistance it requires. By evaluating repair trajectories progressive metrics rather than only final pass/fail outcomes, \tool provides a finer-grained assessment of LLM code-improvement capability.
\end{abstract}


\section{Introduction}
\label{sec:intro}

Large language models (LLMs) have demonstrated strong capabilities across diverse software engineering (SE) tasks, ranging from code generation~\cite{DBLP:journals/corr/abs-2107-03374,jain2025livecodebench} to automated program repair (APR)~\cite{jimenez2024swebench,zhang2024autocoderover,xia2024agentlessdemystifyingllmbasedsoftware}. In code generation, models are presented with a natural language problem statement and evaluated on whether the generated program is functionally correct. In the case of APR, they are evaluated on whether the generated patch for a given buggy program restores correctness without introducing regressions. Despite their apparent differences in problem formulation, both settings share the same success measurement criterion: a binary, instance-level pass/fail outcome determined by the associated test suite.

While a good first-approximation, such binary measures are \textit{sparse} and 
provide a narrow view of overall performance. A model is considered \textit{successful} only if its generated solution passes all tests in the associated test suite. In practice, however, test suites are often weak and rarely exhaustive~\cite{DBLP:conf/sigsoft/SmithBGB15,DBLP:journals/corr/abs-2503-15223}. Thus, a model-generated solution that fixes the reported failure but introduces subtle regressions may still be counted as correct, leading to an inflated assessment of model performance~\cite{openai_swebench_verified_2024}. 
Even when a model is \textit{unsuccessful}, pass/fail outcomes are uninformative: a solution that fails one edge case is treated the same as one that fails all tests, nor does a partially correct solution receive any credit for progress.

The limitations of binary measures are further exposed in workflows where models operate in {\em iterative refinement} loops, \textit{progressively improving} upon an initial solution through compiler feedback~\cite{DBLP:conf/icml/YasunagaL20,dai2025feedbackeval}, execution information~\cite{DBLP:conf/acl/ZhangLLLJ23,DBLP:conf/iclr/ChenLSZ24,jain2025livecodebench}, or model-generated diagnostics~\cite{DBLP:journals/corr/abs-2502-02928,dai2025feedbackeval}. We refer to this process of transforming an incorrect or incomplete program into a more correct one through feedback-guided refinement as {\bf \em code improvement}. Despite its inherently incremental nature, evaluation still remains limited to the binary outcome of the final solution, ignoring the refinement trajectory altogether. 
Such an assessment can therefore be misleading: a stronger model may produce an initial patch that fails the current test suite but moves the program substantially closer to the correct behavior, while a weaker model may pass the available tests by simply making a shallow edit. Under a binary pass-rate metric, the weaker model would be deemed superior, though the stronger model demonstrates more effective use of feedback and better generalization. Accordingly, we position that \textit{benchmarking a model's ability to make progress yields a more fine-grained and faithful measure of its real-world utility}. 

However, measuring progress under code improvement is not straightforward, as model performance is highly sensitive to the informativeness of the feedback provided~\cite{han2025convcodeworld,pan2025benchmarkstalk}. For instance, a hint that directly reveals the solution provides little evidence of model reasoning, while one that is too abstract to act upon conflates feedback quality with model failure. Therefore, the feedback provided to guide code improvement must itself be model-dependent and calibrated to model capabilities. In particular, a faithful progress-centric evaluation should consider the trajectory of code improvement: 
\textit{informativeness of feedback provided at each intermediate step, how much corresponding progress is made, and whether improvement remains monotonic across iterations}? 

To address these challenges, we introduce \tool, a \textit{\textbf{progressive and adaptive benchmark}} for evaluating \textbf{\textit{code improvement}}. The central mechanism of \tool is \textit{progressive hinting}, a structured feedback protocol that controls hint difficulty along two dimensions. The first, \textit{failure-region control} determines \textit{what} the hint is about. It groups hidden failing tests into failure scenarios, where each scenario corresponds to a region of incorrect program behavior, such as a shared edge case, input pattern, or execution path. By prioritizing which failing tests are exposed to the {\em candidate model},
\tool via its feedback model steers the candidate model's fixing toward specific regions of incorrect behavior. 
The benchmark can progressively provide hints that expose different failure regions and measure how the model improves within the hidden failure space. 
The second, \textit{hint-depth control} determines \textit{how much} repair-relevant information is revealed. Inspired by Item Response Theory (IRT), it starts from coarse-grained symptoms to fine-grained implementation-level feedback. Together, both dimensions define failure sub-regions selected relative to the capabilities of the model under evaluation. 
Such within-instance adaptability ensures that the trajectory of code improvement better reflects model capability rather than the informativeness of an unstructured feedback~oracle.

By design, progressive hinting governs \textit{how} a model is~allowed to improve. Accordingly, for each instance, we measure progress at each intermediate step, awarding a \textit{partial credit} that reflects the proportion and significance of failures resolved rather than treating every step as a binary success or failure. For this, we have metrics measuring the progress
where outcomes are continuous rather than binary. 
{\tool} thus achieves both progressivity and adaptivity, yielding a benchmark that is more faithful to a model's true capabilities. In this paper, we instantiate code improvement as iterative automated program repair (APR). We use ``repair'' to refer to each concrete code revision within this benchmark, while “code improvement” denotes the broader evaluation task.



In this paper, we make the following contributions:

\textbf{(1) {\tool}: A new progressive and adaptive benchmark paradigm}. Data and code is available at~\cite{pairbench-website}.


\textbf{(2) A suite of progress-centric metrics} that measure different aspects of interactive code improvement.

\textbf{(3) A controlled interactive feedback protocol.} We propose \emph{progressive hinting}, a structured feedback protocol that jointly controls \emph{failure regions} and \emph{hint depth}.

\textbf{(4) Empirical study.} We instantiate \tool on~interactive code repair and evaluate multiple state-of-the-art LLMs.

\section{Motivation and Key Ideas}
\label{sec:motiv}

    





\subsection{Binary Pass/Fail Outcomes Mask Improvement Ability}

\begin{figure}[t]
\centering
\includegraphics[width=0.95\linewidth]{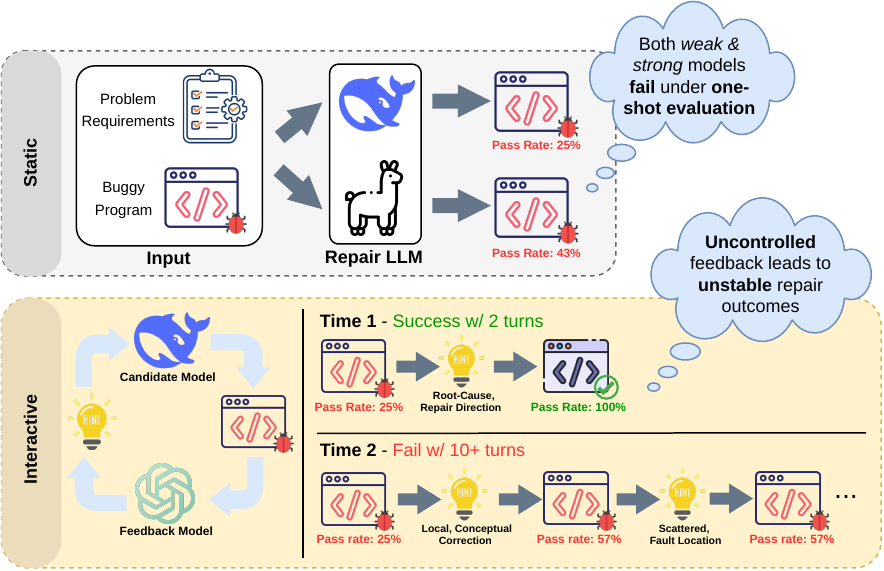}
\vspace{-6pt}
\caption{Motivation for progress-centric interactive repair evaluation. Static pass/fail evaluation can hide partial improvement, while uncontrolled feedback can make interactive repair outcomes unstable across runs.}
\label{fig:motivation}
\end{figure}

Code-generation and APR benchmarks often evaluate models through final functional correctness: a generated code or patch is executed against a test suite and marked as either passed or failed. However, this strategy compresses model capability into a single binary outcome. In a one-shot evaluation, a weaker model may achieve a higher hidden-test pass rate than a stronger one, making it appear better~under a standard pass-rate metric (Fig.~\ref{fig:motivation}). However, after receiving targeted feedback, the stronger model may repair the solution while the weaker model still fails. 
Pass@k only partially addresses this. Although it samples multiple attempts, each attempt is still evaluated as an {\em independent one-shot generation}.
Thus, one-shot correctness can misrepresent \emph{code-improvement ability}, which depends not only on how many tests a model initially passes, but also on whether it uses feedback to correct the underlying failure. We call an evaluation {\bf \em progress-centric} if it measures not only final correctness, but also the sequence of improvements made during refinement: which failures are fixed, whether already-correct behavior is preserved in a monotonic progress, and how much assistance is needed. 



\vspace{-6pt}
\subsection{Interactive Feedback Helps, but Must Be Controlled}

A natural way to address the limitation of one-shot evaluation is to make the benchmark interactive. Instead of judging only the first candidate solution, an interactive benchmark allows the model to receive feedback, and revise and improve its code. This better matches {\em practical LLM-based interactive refinement} workflows.
Recent benchmarks have started to move in this direction as they provide execution outputs, compiler/test diagnostics, or natural-language feedback (Section~\ref{sec:compare}).
Those existing interactive feedback benchmarks do not fully address the requirements of progress-centric evaluation. A faithful evaluation of code improvement should not only ask whether feedback eventually helps, but should also characterize the repair trajectory as defined earlier.
In existing settings, there are two uncontrolled factors on the feedback:
1) {\bf \em the underlying failure being targeted} or 2) {\bf \em the amount of repair-relevant information being revealed}.


This limitation may lead to inconsistent evaluation across runs, even with the same model on the same problem. Different feedback runs may thus produce different repair trajectories and benchmark scores. In one run, the feedback may quickly identify the decisive failure mode or provide a highly direct hint, allowing the model to repair the solution after only a few turns. In another run, the feedback may focus on a local, scattered, less informative symptom, target a different failing behavior, or remain too vague to guide the repair, causing the same model to fail even after many turns (Fig.~\ref{fig:motivation}). 
In such cases, the benchmark is no longer measuring only the model’s code-improvement ability; it is also measuring the variability and informativeness of the feedback itself.

\subsection{Key Ideas}


The instability illustrated in Fig.~\ref{fig:motivation} suggests that interactive feedback is necessary but insufficient. 
To reduce such instability and support faithful progress-centric evaluation, a benchmark must make feedback comparable across runs and models. We argue that feedback should be controlled along two dimensions: the {\bf \emph{failure region}} being addressed and the {\bf \emph{hint depth}} of the feedback. These two dimensions separate the target of feedback from the strength of feedback.


\subsubsection{\bf \em Failure-region control} This dimension determines which part of the hidden failure space the feedback targets. Instead of generating feedback from an arbitrary failing test case, we group hidden failing tests into failure scenarios, where each scenario corresponds to a related region of incorrect behavior.
To make feedback comparable across turns, each hint is anchored to a selected scenario, whose scope can be approximated by the number of covered failing test~cases.

\subsubsection{\bf \em Hint-depth Control} This dimension determines how much repair-relevant information the hint reveals. Lower-depth hints expose only external evidence of the failure, e.g., symptoms or input patterns, while higher-depth hints provide increasingly direct repair guidance, e.g., missing state, suspicious code regions, conceptual corrections, or implementation directions. We use a six-level hint-depth scale (Table~\ref{tab:hint_level}), ranging from symptom-only feedback to implementation direction.


\subsubsection{\bf \em Progressive Evaluation}

A code-improvement benchmark should be \emph{progressive}. A buggy solution may fail for multiple reasons.
If interaction repeatedly focuses on the same unresolved failure, the benchmark may observe many repair attempts without measuring broader improvement. 

The failure-region control enables \emph{progressive} evaluation. The benchmark can progressively provide hints that expose different failure regions and measure how the model improves within the hidden failure space. After each hint, the benchmark can evaluate whether the model fixes the targeted scenario, how much progress the model makes after each hint, whether the repair generalizes to other unhinted failures, and whether previously passing tests remain passing. Progressive evaluation thus treats repair as a trajectory over the hidden failure space. 



\subsubsection{\bf \em Adaptive Evaluation}

A code-improvement benchmark should also be {\em adaptive}. Even when two hints target the same failure region, they may reveal very different amounts of repair-relevant information. A shallow hint may describe only the observed symptom, while a more direct hint may reveal the input pattern, missing state, suspicious code region, conceptual correction, or concrete repair direction. 

Hint-depth control enables \emph{adaptive} evaluation. 
Our benchmark treats hint depth as a way to calibrate feedback difficulty to the model's demonstrated code-improvement capability. A model that repairs a failure from shallow evidence should receive less direct guidance, while a model that remains stuck can be given deeper hints that reveal more repair-relevant information. This allows the benchmark to estimate not only whether a model eventually fixes the program, but also how much assistance it requires to make progress. 


\section{Related Work}
\label{sec:compare}

\newcommand{\Yes}{\ding{51}}
\newcommand{\No}{\ding{55}}
\newcommand{\Partial}{\ensuremath{\triangle}}

\newcolumntype{L}[1]{>{\raggedright\arraybackslash}m{#1}}
\newcolumntype{C}[1]{>{\centering\arraybackslash}m{#1}}

\begin{table}[t]
\centering
\scriptsize
\renewcommand{\arraystretch}{1.16}
\setlength{\tabcolsep}{1.7pt}

\caption{Comparison with representative evaluation settings.}
\label{tab:benchmark_comparison}

\resizebox{0.49\textwidth}{!}{
\begin{tabular}{L{2.07cm} C{0.85cm} C{1.15cm} C{0.8cm} C{1.3cm} C{1.2cm} C{0.95cm}}
\toprule
\multirow{2}{*}{\shortstack{\textbf{Evaluation}\\\textbf{setting}}}
& \textbf{Exec.}
& \textbf{Multi-turn}
& \textbf{NL}
& \textbf{Hidden-test}
& \textbf{Progressive}
& \textbf{Adaptive} \\
& \textbf{grounded}
& \textbf{repair}
& \textbf{feedback}
& \textbf{protected}
& \textbf{evaluation}
& \textbf{feedback} \\
\midrule

Static code generation~\cite{DBLP:journals/corr/abs-2107-03374,austin2021program,jain2025livecodebench}
& \Yes & \No & \No & \Yes & \No & \No \\

Static program repair~\cite{jimenez2024swebench}
& \Yes & \No & \No & \Yes & \No & \No \\

Interactive execution feedback~\cite{yang2023intercode}
& \Yes & \Yes & \No & \No & \Partial & \No \\

Feedback-driven repair~\cite{dai2025feedbackeval}
& \Yes & \Yes & \Yes & \Partial & \Partial & \No \\

Conversational feedback~\cite{han2025convcodeworld,pan2025benchmarkstalk}
& \Partial & \Yes & \Yes & \Partial & \Partial & \Partial \\

\textbf{\tool}
& \Yes & \Yes & \Yes & \Yes & \Yes & \Yes \\

\bottomrule
\end{tabular}
}

\vspace{0.3em}
{\footnotesize
\textbf{Legend:}
\Yes~= supported;
\No~= not supported;
\Partial~= partially supported.
}
\end{table}

We compare existing code-evaluation benchmarks along six dimensions: execution grounding, multi-turn repair, natural-language feedback, hidden-test protection, progressive evaluation, and adaptive feedback. Execution grounding means that generated code is executed and evaluated against tests or an interactive environment. Multi-turn repair means that the model can revise its solution after receiving feedback. Natural-language feedback means that the benchmark can provide verbal guidance beyond raw execution output. Hidden-test protection captures whether the candidate model is shielded from hidden test cases, expected outputs, and raw hidden execution traces. This is crucial because exposing hidden~failures makes repair easier but turns the task into test-specific patching. In realistic settings, the model only sees the problem requirements and public tests, while hidden tests are for~evaluation.


\begin{figure*}[t]
\centering
\includegraphics[width=0.9\linewidth]{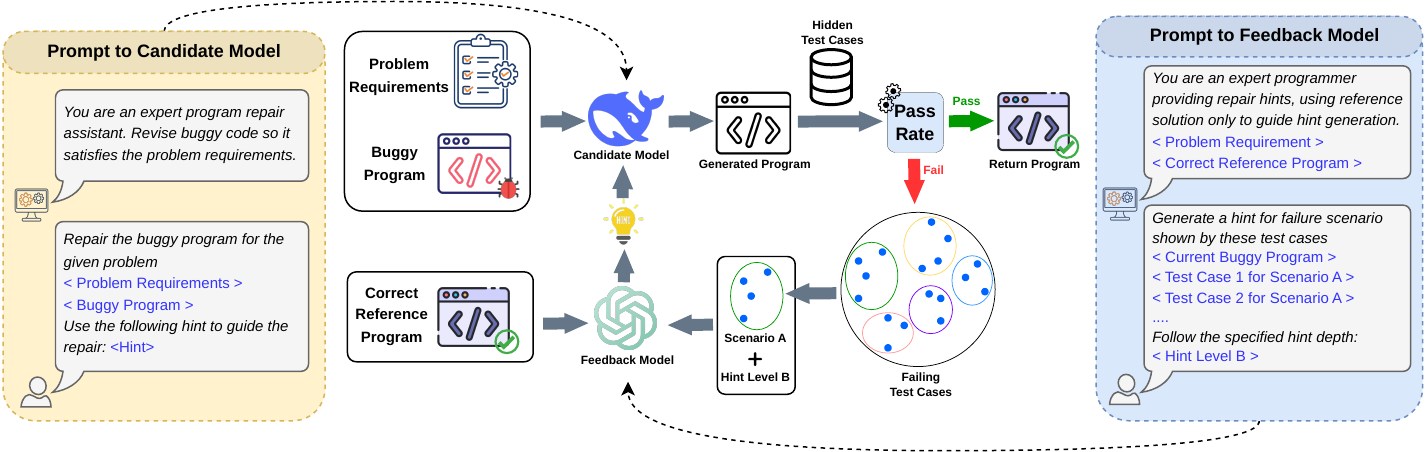}
\vspace{-6pt}
\caption{Overview of \tool. Hidden tests are grouped into failure scenarios, a feedback model generates level-controlled hints, and the candidate model iteratively repairs the buggy program.}
\label{fig:pairbench}
\end{figure*}

Table~\ref{tab:benchmark_comparison} summarizes how \tool differs from representative benchmarks. Static code-generation and APR benchmarks evaluate if a model can synthesize a correct program from a problem statement or fix a buggy code~\cite{DBLP:journals/corr/abs-2107-03374,austin2021program,hendrycks2021measuring,evalplus,evalperf,10.1145/2610384.2628055,10.1145/3368089.3417943,10.1145/3135932.3135941}.
LiveCodeBench includes self-repair and execution-related tasks~\cite{jain2025livecodebench}. SWE-bench~\cite{jimenez2024swebench} and SWE-bench Verified~\cite{openai_swebench_verified_2024} move evaluation toward realistic repo-level repair by asking models to resolve real GitHub issues.
Another line of work evaluates models in settings where code execution results are returned to the model. InterCode treats code as actions and execution results as observations~\cite{yang2023intercode}. Other work on self-editing, self-debugging, and execution-guided code generation studies whether models can use compiler messages, runtime errors, failed tests, or execution information to revise their programs~\cite{DBLP:conf/acl/ZhangLLLJ23,DBLP:conf/iclr/ChenLSZ24,DBLP:journals/corr/abs-2502-02928}.
Recently, FeedbackEval studies how LLMs use different feedback types in code repair~\cite{dai2025feedbackeval}. Conversational benchmarks, e.g., ConvCodeWorld and When-Benchmarks-Talk, show that model behavior can change substantially depending on the feedback setting~\cite{han2025convcodeworld,pan2025benchmarkstalk}.

The comparison shows that the missing capability is not simply interaction, but controlled interaction.

\section{\textbf{{\tool}}: \textbf{P}rogressive and \textbf{A}daptive \textbf{I}nteractive \textbf{R}efinement Benchmark}
\label{sec:bench}


Based on the above requirements, {\tool} operationalizes progress-centric evaluation via the pipeline shown in Fig.~\ref{fig:pairbench}. The benchmark consists of four components:

(1) a scenario constructor that groups currently failing hidden tests into failure regions;

(2) a hint generator that produces feedback at a specified hint depth for a selected scenario;

(3) an adaptive policy that selects the next scenario and hint level based on observed repair behavior; and

(4) a trajectory evaluator measuring targeted repair, broader repair, preservation, hint efficiency, and final convergence. 

An instance in our benchmark is defined as $\langle x_i, c_{i,0}, \mathcal{T}*{i} \rangle$, where $x_i$ is the problem statement, $c*{i,0}$ is the buggy solution, and $\mathcal{T}_{i}$ is the hidden test suite.


Given an instance, the {\em candidate model} first produces an initial repaired program from the problem requirements and buggy program (Fig.~\ref{fig:pairbench}). Hidden tests are hidden from the candidate model but available to the benchmark controller, which evaluates the generated program on the hidden test suite, records the pass rate, and extracts the remaining failing test cases. These failing tests are grouped into a selected failure scenario S, and the controller assigns a hint depth level H.

The {\em feedback model} then receives the problem requirement, the correct reference program, the current buggy program, the selected scenario test cases, and the requested hint level. It generates only a controlled hint for that scenario and depth. 

If the program still fails at least one hidden test case, {\tool} enters a multi-turn refinement loop. At each turn, the benchmark selects a controlled failure scenario, asks the feedback model to generate a hint at the specified depth, asks the candidate model to revise the code, and re-evaluates the revised program on the hidden test cases. At turn (t), the candidate model receives:
$
\langle x_i, c_{t-1}, h_t, \mathcal{H}*{<t} \rangle
$
where \(x_i\) is the problem statement, \(c_{t-1}\) is the current code, \(h_t\) is the controlled hint, and \(\mathcal{H}_{<t}\) is the prior interaction history. It is {\bf not} given hidden tests, oracle traces/outputs, or reference~code.

\vspace{-4pt}
\subsection{Scenario Constructor: Scenario-Level Failure Modeling}
\label{subsec:failure-modeling}

Let $\mathcal{T}_i=\{1,\ldots,n_i\}$ denote the hidden test cases for instance program $i$. After turn $t$, let $F_{i,t}\subseteq \mathcal{T}_i$ be the set of failing test cases and $P_{i,t}=\mathcal{T}_i\setminus F_{i,t}$ be the passing ones. {\tool} groups the currently failing test cases into \emph{failure regions}, where each region corresponds to a related behavior that the current program handles incorrectly.

For each failing test case, {\tool} constructs a failure signature using three signals: 1) the reference execution trace, 2) the expected-output shape, and 3) the candidate failure type. The {\em reference trace} records the source lines executed by the correct solution, approximating the behavioral path exercised by the test case. The {\em output-shape} signature captures coarse output structure, such as a single token, a binary line, a grid, or a multi-line output. The {\em candidate failure type} distinguishes different failure types including timeouts, runtime errors, wrong token counts, line counts, or values.~Failing test cases with the same signature are grouped into one scenario:
\vspace{-6pt}
\[
G_{i,t}^{(k)} = \{j \in F_{i,t} : \sigma(j)=k\},
\]
where $\sigma(j)$ is the failure signature of the test case $j$.

The expected-output shape is computed from the oracle output and captures the structural form of the required answer, while the candidate failure type is computed from the comparison between the candidate output and the oracle output.

To avoid over-fragmenting the failure space, we use a~hierarchical grouping strategy. 
We group failing tests into {\em failure scenarios}, each representing a {\em region} of incorrect behavior.
It uses the full signature consisting of trace, output shape, and failure type. 
If more than \(K_{\max}\) scenarios are produced or if the median scenario size falls below \(s_{\min}\), it backs off to output-shape plus failure type, and finally to failure type alone.
\(K_{\max}\) and \(s_{\min}\) are predefined.
The benchmark prioritizes the active failure region by {\em the number of current failing test cases it covers}, using trace coverage~as a secondary tiebreaker. Thus, each hint is anchored to a concrete and reproducible region of hidden failure behavior. The benchmark {\em progressively} provides hints that expose more failure regions and measure how the candidate model improves within the failure space. 
We do not claim that the signature perfectly identifies semantic bug causes. Rather, it provides a reproducible approximation of execution-grounded failure regions. The grouping is used to control feedback targets, not to infer oracle root causes.

\vspace{-4pt}
\subsection{Scaffolded Hint-Depth Design}
\label{subsec:hint-depth}



According to the foundation on formative feedback and intelligent tutoring systems, effective feedback should be task-focused, specific enough to support progress~\cite{hattie2007power,shute2008formative}. In educational scaffolding, assistance is adjusted to the learner's current capability and gradually increased or faded as needed~\cite{wood1976tutoring,vygotsky1978mind}. {\tool} applies this idea to interactive code improvement: a model should first be given weak evidence about the failure, and only receive more direct hinting information when weaker feedback is insufficient. 


\begin{table}[t]
\centering
\footnotesize
\caption{Hint-depth scale for controlled feedback. Higher levels reveal more direct repair information.}
\label{tab:hint_level}
\vspace{-6pt}
\setlength{\tabcolsep}{1.5pt}
\resizebox{0.98\columnwidth}{!}{%
\begin{tabular}{ll}
\toprule
\textbf{Hint Level} & \textbf{Information Revealed and Example} \\
\midrule

L1. Symptom
& Only the observed incorrect behavior; no input \\
& pattern, cause, or code location.\\
& Example: ``The output is larger than expected,\\
& suggesting that something may be counted extra.'' \\

L2. Input pattern
& The type of input or edge case exposing the failure.\\
& Example: ``This failure occurs when the input\\
& contains duplicate values.'' \\

L3. State tracking
& The information that the solution fails to preserve,\\
& update, or track correctly.\\
& Example: ``The solution does not keep enough\\
& information about repeated values across iterations.'' \\

L4. Fault location
& The suspicious region of the candidate implementation.\\
& Example: ``The issue is likely in the loop that updates\\
& the count after reading each value.'' \\

L5. Concept. correction
& The missing invariant, condition, or reasoning needed\\
& to handle the scenario.\\
& Example: ``The solution needs to distinguish between\\
& seeing a value once and seeing it multiple times.'' \\

L6. Repair direction
& A concrete implementation direction, without giving\\
& the full corrected code.\\
& Example: ``Revise the update logic so repeated values\\
& are accumulated rather than overwritten.'' \\

\bottomrule
\end{tabular}%
}
\end{table}

We organize hints along an increasing informativeness axis, using the six-level scale shown in Table~\ref{tab:hint_level}. The levels progress from {\em outcome-level feedback to repair-level guidance}.~L1 describes only the observable symptom, corresponding to~minimal feedback. L2 reveals the input condition under which the failure occurs, making the feedback diagnostic but still implementation-agnostic. L3 identifies the missing state or condition the solution fails to track, moving from external~behavior to internal reasoning. L4 localizes the suspicious part of the candidate implementation, similar to context-specific hints in intelligent tutoring systems. L5 states the missing~invariant or conceptual correction, and L6 gives a concrete repair direction without giving code, pseudocode, or exact patch.

This hierarchy separates \emph{diagnostic evidence} from \emph{repair disclosure} and makes hint depth a controlled assistance variable. Lower levels test whether the model can infer the bug from observed behavior, while higher levels reveal increasingly direct guidance when weaker feedback is insufficient. Thus, repairing a scenario at a lower level indicates stronger diagnostic and refinement ability, whereas requiring higher-level hints indicates greater dependence on explicit guidance. 

This design also supports adaptive evaluation: after a successful repair, the benchmark can reduce the starting hint level for the next scenario, analogous to fading support after demonstrated competence; if the model remains stuck, it escalates the hint level toward the model's effective capability. In this way, the hint scale operationalizes scaffolded interaction for APR while keeping feedback comparable across models and runs.



\subsection{Controlled Hint Generator}

To preserve the meaning of the hint-depth scale, we~constrain hint generation through level-specific prompt templates, evidence masks, and leakage validation. Each level~is associated with an allowed information boundary that decid\-es~what~evidence the feedback model may observe and~what information the generated hint may reveal. E.g., an L2 hint may use failing-input features and observed output mismat\-ches, but the feedback model is not given candidate-code locations, suspicious code regions, or repair-relevant~invariants. In contrast, an L4 hint may additionally receive candidate-code context since fault localization is allowed at that level.

After generation, {\tool} validates each hint against level-specific leakage criteria. A hint is rejected if it reveals information reserved for a higher hint level, such as exposing a suspicious code location in an L2 hint or suggesting a concrete repair direction in an L4 hint. Rejected hints are regenerated until they satisfy the requested information boundary or the budget is exhausted. This helps ensure that hint depth functions as a controlled assistance rather than merely a post-hoc label.



\subsection{Progressive and Adaptive Hint Policy}
\label{subsec:hint-policy}

{\tool} uses a progressive policy over failure regions and adaptive over hint depth (Algorithm~\ref{alg:pairbench}).

{\color{custom-blue}
\subsubsection{Progressive Policy} 
Generally, {\tool} selects a failure scenario, provides feedback for that scenario, evaluates whether the model fixes it, and moves to another unresolved scenario.
At turn $t$, it re-evaluates the current program and groups the remaining failing tests into scenarios. Next, it must decide which failure region to address. We prioritize scenarios by their current failing-test coverage, since a scenario covering more failing tests represents a larger unresolved portion of the hidden failure space. This choice makes each hint target the largest remaining opportunity for measurable improvement, while completed and deferred scenarios are excluded to ensure that the interaction progresses across distinct failure regions. Formally, let $\mathcal{U}_{i,t}$ denote the set of unresolved selectable scenarios at turn $t$. The active scenario is selected as follows:
\[
G_{i,t}
=
\arg\max_{G\in \mathcal{U}_{i,t}}
|G|
\]
Ties are broken using trace coverage and then a deterministic scenario identifier. The benchmark then samples up to $m$ representative failing test cases from $G^{*}_{i,t}$ to ground the hint. These sampled tests are used only for hint generation; scenario completion is evaluated over the full selected scenario. If all tests in $G^{*}_{i,t}$ pass after revision, the scenario is marked as completed. Otherwise, the benchmark continues on the same scenario until either the maximum hint depth or the per-scenario turn limit is reached.
A completed scenario is re-opened if any of its tests fail after a later revision, ensuring that regressions remain part of the active failure space.
}




To avoid repeatedly prioritizing large scenarios, {\tool} excludes completed and deferred scenarios and enforces a per-scenario turn budget. This ensures that the interaction progresses across distinct failure regions rather than repeatedly focusing on one dominant failure mode.
We prioritize coverage because it maximizes measurable repair opportunity at each turn, while the scenario-turn limit prevents the policy from spending the entire budget on a single unresolved region.

\subsubsection{Adaptive Policy} The policy is adaptive because the next hint level depends on the model's observed repair behavior. 
{\color{custom-blue}Inspired by adaptive testing and IRT-style assessment, {\tool} treats hint depth as an ordered assistance scale with a deterministic staircase policy as a lightweight operationalization: success decreases the next starting hint level, while failure increases it. This policy does not require fitting IRT parameters, but preserves the central idea of calibrating item assistance to observed ability.}
Specifically, each scenario begins with the current starting level, initially L1. If the model repairs the selected tests, {\tool} treats the model as capable of using the current level of feedback and starts the next scenario with a weaker or equal hint:
$
\ell_{t+1}=\max(1,\ell_t-1).
$
If the selected test cases still fail, {\tool} increases the hint depth:
$
\ell_{t+1}=\min(6,\ell_t+1).
$
The interaction stops when all hidden tests pass, no remaining scenario can be selected, or the max turn budget is~reached. In brief, successful repair leads to faded assistance, while failure leads to more direct~guidance. 



\begin{algorithm}[t]
\small
\caption{Progressive and adaptive hint policy algorithm}
\label{alg:pairbench}
\begin{algorithmic}[1]
\Require hidden tests $\mathcal{T}$, program $c_0$, turn budget $T$, max hint= $6$
\State Evaluate $c_0$ and obtain failing set $F_0$
\State $\ell \gets 1$, completed $\gets \emptyset$, deferred $\gets \emptyset$
\For{$t=1$ to $T$}
    \If{$F_{t-1}=\emptyset$}
        \State \textbf{break}
    \EndIf
    \State Group $F_{t-1}$ into failure scenarios $\{G^{(k)}\}$
    \State Select unresolved scenario $G_t$ with largest remaining fail cov.
    \State Select representative failing tests $S_t\subseteq G_t$
    \State Generate a level-$\ell$ hint grounded in $S_t$
    \State Candidate revises $c_{t-1}$ into $c_t$
    \State Evaluate $c_t$ and obtain $F_t$
    \If{$G_t\cap F_t=\emptyset$} 
        \State completed $\gets$ completed $\cup \{G_t\}$
        \State $\ell \gets \max(1,\ell-1)$
    \ElsIf{$\ell=6$ or scenario-turn limit is reached}
        \State deferred $\gets$ deferred $\cup \{G_t\}$
    \Else
        \State $\ell \gets \ell+1$
    \EndIf
\EndFor
\end{algorithmic}
\end{algorithm}

\subsection{Trajectory-Level Evaluator and Evaluation Metrics}
\label{subsec:metrics}


With the same notations in Section~\ref{subsec:failure-modeling}, we define our~metrics.
Turn $t=0$ is the initial repair before any hint. The~pass rate (PR) is defined as $p_{i,t}=\frac{|P_{i,t}|}{|\mathcal{T}_i|}$. Let $\mathcal{I}$ be the benchmark instances. At repair turn $t$, let $G_{i,t}$ $\subseteq F_{i,t-1}$ be the failure scenario targeted by the hint, and let $b_i$=$\max_{0\leq t\leq B}p_{i,t}$ be the {\em best pass rate} in the turn budget $B$.

\paragraph{Initial Fix Rate}
This measures unaided repair ability before feedback:
$
\mathrm{InitialFix}
=
\frac{1}{|\mathcal{I}|}
\sum_{i\in\mathcal{I}}
\mathbf{1}[p_{i,0}=1].
$
It separates one-shot repair capability from interactive improvement.

\paragraph{Targeted Repair Success (TRS)}
This measures if a model fixes the failure region described by the current hint:
\[
\mathrm{TRS}_{i,t}
=
\frac{|G_{i,t}\cap P_{i,t}|}{|G_{i,t}|}.
\]
It captures the precise use of controlled feedback. 

\paragraph{Broader Repair Gain (BRG)}
This measures progress beyond the hinted scenario:
$
\mathrm{BRG}_{i,t}
=
\frac{|(F_{i,t-1}\setminus G_{i,t})\cap P_{i,t}|}{|\mathcal{T}_i|}.
$
It captures if a model generalizes a targeted hint to other~hidden failures. We normalize by the total number of hidden tests to reflect absolute contribution to overall pass-rate improvement. 
 
\paragraph{Behavior Preservation (BP)}
This measures whether a revision preserves behavior that was already correct:
\[
\mathrm{BP}_{i,t}
=
1-
\frac{|P_{i,t-1}\cap F_{i,t}|}{|\mathcal{T}_i|}.
\]
It penalizes newly introduced regressions. We use the same total-test normalization so the regression penalty is comparable across instances and directly aligned with pass-rate change.

\paragraph{Progress Monotonicity Rate (PMR)} This measures whether the repair trajectory improves monotonically across interaction turns. Since $p_{i,t}$ denotes the hidden-test pass rate after turn $t$, PMR compares each repaired program with the program from the previous turn. Let $\tau_i\leq B$ be the last evaluated repair turn for instance $i$. We define: 
\[ \mathrm{PMR}_i = 
\frac{1}{\tau_i}
\sum_{t=1}^{\tau_i}
\mathbf{1}[p_{i,t}\geq p_{i,t-1}].
\]
PMR measures the fraction of refinement steps that preserve or improve the previous pass rate, with higher values indicating a more stable trajectory with fewer regressions.

\paragraph{Hint Efficiency (HE)}
This measures how much aid a model needs to close targeted failure scenarios. For each attempted scenario $g$, let $\ell_{i,g}\in\{1,...,6\}$ be the hint level at which the scenario is closed. We assign a scenario-level score
\[
h_{i,g}
=
\begin{cases}
7-\ell_{i,g}, & \text{if scenario } g \text{ is fully repaired},\\
0, & \text{otherwise}.
\end{cases}
\]
The instance-level Hint Efficiency is the average over all attempted scenarios $\mathcal{G}_i$:
$
\mathrm{HE}_i
=
\frac{1}{|\mathcal{G}_i|}
\sum_{g\in\mathcal{G}_i} h_{i,g}.
$
Higher HE indicates that a model closes failure scenarios with weaker, less explicit hints; unresolved scenarios get no credit.

\paragraph{Gap Closure}
This measures the normalized fraction of remaining repair opportunity that the model closes:
$
\mathrm{GC}_i=\frac{b_i-p_{i,0}}{1-p_{i,0}}.
$
It gives credit for progress.

\paragraph{Final Fix Rate}
This measures complete repair within the interaction budget:
$
\mathrm{FinalFix}
=
\frac{1}{|\mathcal{I}|}
\sum_{i\in\mathcal{I}}
\mathbf{1}[\exists t\leq T,\;p_{i,t}=1].
$
It preserves the standard end-to-end correctness criterion.

\paragraph{Turns to Fix}
This measures convergence speed among initial failing instances that are eventually fixed. Let
$
\mathcal{S}
=
\{i\in\mathcal{I}:p_{i,0}<1 \wedge \exists t\leq T,\;p_{i,t}=1\}.
$
Then
\[
\mathrm{TurnsToFix}
=
\frac{1}{|\mathcal{S}|}
\sum_{i\in\mathcal{S}}
\min\{t:p_{i,t}=1\}.
\]
It distinguishes models that reach the same final success rate but require different numbers of refinement turns. 

Turn-level metrics are averaged within each initially failing instance and then across initially failing instances. Initial Fix Rate and Final Fix Rate are computed over all instances. 

\newcommand{\goodthree}[1]{\textbf{\textcolor{green!50!black}{#1}}}
\newcommand{\badthree}[1]{\textbf{\textcolor{red!70!black}{#1}}}

\newcommand{\ffgain}[1]{%
  \kern0.10em%
  \raisebox{1ex}{%
    \scalebox{0.55}{\textnormal{\textcolor{black!75}{+#1}}}%
  }%
}

\begin{table*}[!htpb]
\centering
\scriptsize
\setlength{\tabcolsep}{3pt}
\renewcommand{\arraystretch}{1.18}
\caption{Progress-centric leaderboard for controlled interactive program repair. \textcolor{green!50!black}{Green} marks the top-three models and \textcolor{red!70!black}{red} marks the bottom-three models for each metric, following the metric direction (RQ1).}
\label{tab:rq1_leaderboard}
\resizebox{\textwidth}{!}{
\begin{tabular}{l c c c c c c c c c}
\toprule
\multirow{3}{*}{\textbf{Model}} &
\multicolumn{1}{c}{\textbf{Before Feedback}} &
\multicolumn{5}{c}{\textbf{Interactive Repair}} &
\multicolumn{3}{c}{\textbf{After Feedback / Outcome}} \\
\cmidrule(lr){2-2} \cmidrule(lr){3-7} \cmidrule(lr){8-10}

& \multirow{2}{*}{\textbf{Initial Fix(\%)$\uparrow$}}
& \textbf{Targeted}
& \textbf{Broader}
& \textbf{Behavior}
& \textbf{Progress}
& \textbf{Hint}
& \textbf{Final}
& \textbf{Gap}
& \textbf{Turns} \\

&
& \textbf{Repair(\%)$\uparrow$}
& \textbf{Repair(\%)$\uparrow$}
& \textbf{Preservation(\%)$\uparrow$}
& \textbf{Monotonicity(\%)$\uparrow$}
& \textbf{Efficiency(0--6)$\uparrow$}
& \textbf{ Fix(\%)$\uparrow$}
& \textbf{Closure(\%)$\uparrow$}
& \textbf{to Fix$\downarrow$} \\
\midrule

DeepSeek V3.2
& \goodthree{65.68}
& \goodthree{71.59}
& \goodthree{10.19}
& \goodthree{98.30}
& \goodthree{90.66}
& \goodthree{4.79}
& \goodthree{99.31}\ffgain{33.63}
& \goodthree{98.80}
& \goodthree{2.54} \\

Gemini 2.5 Flash Lite
& \goodthree{53.18}
& \goodthree{59.00}
& \goodthree{10.97}
& 93.65
& \goodthree{81.39}
& \goodthree{3.76}
& \goodthree{95.45}\ffgain{42.27}
& \goodthree{97.51}
& \goodthree{3.27} \\

Qwen3 Coder 30B A3B Instruct
& \goodthree{35.90}
& 55.89
& 8.42
& \goodthree{95.21}
& \goodthree{82.19}
& 3.61
& \goodthree{90.90}\ffgain{55.00}
& \goodthree{94.25}
& 3.81 \\

GPT-4o-mini
& 29.77
& \goodthree{56.14}
& \badthree{7.14}
& \goodthree{94.92}
& 76.44
& \goodthree{3.80}
& \badthree{85.45}\ffgain{55.68}
& \badthree{89.08}
& \goodthree{3.67} \\

Llama 3.3 70B Instruct
& 27.72
& 53.50
& \badthree{7.59}
& 93.74
& \badthree{71.61}
& \badthree{3.21}
& 85.90\ffgain{58.18}
& 89.32
& \badthree{4.39} \\

Ministral 3 14B
& \badthree{22.95}
& \badthree{49.71}
& \goodthree{10.29}
& \badthree{91.25}
& \badthree{73.17}
& 3.29
& 85.90\ffgain{62.95}
& 91.73
& \badthree{4.33} \\

Mistral Small 3.2 24B
& \badthree{13.40}
& \badthree{49.31}
& 8.05
& \badthree{92.74}
& \badthree{76.35}
& \badthree{3.02}
& \badthree{79.77}\ffgain{66.37}
& \badthree{87.92}
& \badthree{4.32} \\

Gemma 3 27B
& \badthree{10.22}
& \badthree{42.56}
& \badthree{7.84}
& \badthree{90.18}
& 81.24
& \badthree{2.76}
& \badthree{72.04}\ffgain{61.82}
& \badthree{80.06}
& 4.31 \\

\bottomrule
\end{tabular}
}
\end{table*}

\section{Empirical Evaluation}
\label{sec:exp}

For evaluation, we seek to answer the following questions:

\noindent\textbf{RQ1. [Interactive Repair Performance]} 
How effectively do LLMs repair buggy code under controlled feedback, and what aspects of the repair trajectory distinguish stronger models? 

\noindent\textbf{RQ2. [Robustness]} 
Does Controlled Feedback produce more stable interactive repair results than Vanilla Feedback?

\noindent\textbf{RQ3. [Hyperparameter Sensitivity]} 
Do model rankings remain consistent when key benchmark parameters are~varied?


\subsubsection*{Dataset Construction} We collected our dataset from real Codeforces Python submissions in the Hugging Face dataset \cite{matrixstudio_codeforces_python_submissions}. We retain only submissions with the `Wrong Answer' verdict, since these programs compile successfully but fail on execution behavior. We further filter for non-trivial instances by requiring each problem to have more than 100 test cases, an initial pass rate between 10\% and 55\%, 15--80 lines of code, and cyclomatic complexity of at least 8. For each selected buggy submission, we use GPT-OSS 120B to iteratively repair the program with failing test cases as feedback until it passes all tests, producing a validated reference solution that remains close to the original implementation logic. The final dataset contains 440 code-improvement instances, each consisting of {\em a problem statement}, {\em a wrong-answer Python program}, {\em test cases}, and {\em a corresponding reference solution}.

\subsubsection*{Model Selection} We select representative models from diverse model families to cover both closed-source and open-source systems with broadly comparable coding capabilities. Specifically, we chose DeepSeek V3.2, Gemini 2.5 Flash Lite, GPT-4o-mini, Qwen3 Coder 30B A3B Instruct, Llama 3.3 70B Instruct, Ministral 14B, Mistral Small 3.2 24B, and Gemma 3 27B. To keep feedback generation consistent across candidate models, we use GPT-OSS 120B as the feedback model, since it provides strong performance on coding and program-repair tasks. For readability, some tables and figures use the model-family name as shorthand; e.g., `DeepSeek` refers to DeepSeek V3.2, `Gemini` refers to Gemini 2.5 Flash Lite.

\subsubsection*{Evaluation Protocol}
For each instance, the model first attempts a zero-hint repair, which is used to compute Initial Fix Rate. If hidden tests still fail, \tool starts controlled hinting: it selects a failure scenario, provides a hint, asks the model to fix the code, and re-evaluates the result. We~use two parameters: $\alpha$=10 feedback turns per instance and $\beta$=3 turns per failure scenario. Sensitivity is reported in Section~\ref{sec:rq3}.
\subsection{Interactive Repair Performance (RQ1)}

\begin{table*}[t]
\centering
\scriptsize
\setlength{\tabcolsep}{3pt}
\renewcommand{\arraystretch}{1.12}
\caption{Metric-wise ranking summary (RQ1).}
\label{tab:rq1_metric_ranking}
\vspace{-6pt}
\begin{minipage}[t]{0.72\textwidth}
\begin{tabular}{c c c c c c c c c c}
\toprule
\multirow{2}{*}{\textbf{Rank}} &
\textbf{Initial} &
\textbf{Targeted} &
\textbf{Broader} &
\textbf{Behavior} &
\textbf{Progress} &
\textbf{Hint} &
\textbf{Final} &
\textbf{Net} &
\textbf{Turns} \\
&
\textbf{Fix} &
\textbf{Repair} &
\textbf{Repair} &
\textbf{Preservation} &
\textbf{Monotonicity} &
\textbf{Efficiency} &
\textbf{Fix} &
\textbf{Gain} &
\textbf{to Fix} \\
\midrule

\rowcolor{gray!15}
1 & \textbf{DeepSeek} & \textbf{DeepSeek} & \textbf{Gemini} & \textbf{DeepSeek} & \textbf{DeepSeek} & \textbf{DeepSeek} & \textbf{DeepSeek} & \textbf{Gemini} & \textbf{DeepSeek} \\

\rowcolor{gray!15}
2 & \textbf{Gemini} & \textbf{Gemini} & \textbf{Ministral} & \textbf{Qwen} & \textbf{Qwen} & \textbf{GPT} & \textbf{Gemini} & \textbf{Qwen} & \textbf{Gemini} \\

\rowcolor{gray!15}
3 & \textbf{Qwen} & \textbf{GPT} & \textbf{DeepSeek} & \textbf{GPT} & \textbf{GPT} & \textbf{Gemini} & \textbf{Qwen} & \textbf{Mistral} & \textbf{GPT} \\

4 & GPT & Qwen & Qwen & Llama & Gemini & Qwen & Llama & Ministral & Qwen \\

5 & Llama & Llama & Mistral & Gemini & Llama & Ministral & Ministral & Gemma & Gemma \\

6 & Ministral & Ministral & Gemma & Mistral & Ministral & Llama & GPT & DeepSeek & Mistral \\

7 & Mistral & Mistral & Llama & Ministral & Mistral & Mistral & Mistral & Llama & Ministral \\

8 & Gemma & Gemma & GPT & Gemma & Gemma & Gemma & Gemma & GPT & Llama \\

\bottomrule
\end{tabular}%
\end{minipage}
\begin{minipage}[t]{0.18\textwidth}
\centering
\scriptsize
\renewcommand{\arraystretch}{1.12}
\begin{tabular}{lc}
\toprule
\multirow{2}{*}{\textbf{Model}} & \multirow{2}{*}{\shortstack{\textbf{Average} \\ \textbf{Rank}} $\downarrow$} \\
& \\
\midrule
\rowcolor{gray!15} \textbf{DeepSeek} & 1.56\\
\rowcolor{gray!15} \textbf{Gemini} & 2.22\\
\rowcolor{gray!15} \textbf{Qwen} & 2.89\\
GPT & 4.33\\
Ministral & 5.44\\
Llama & 5.78\\
Mistral & 6.33\\
Gemma & 7.44\\
\bottomrule
\end{tabular}
\end{minipage}

\end{table*}

Table~\ref{tab:rq1_leaderboard} reports the main progress-centric results for controlled interactive program repair, while Table~\ref{tab:rq1_metric_ranking} summarizes the metric-wise rankings across the evaluation~dimensions. 

\subsubsection{How much do models benefit from hints?}

Final Fix Rate and Gap Closure (GC) capture complementary aspects of feedback-driven improvement. Final Fix measures whether interaction eventually produces a fully correct program, while GC measures how much of the remaining opportunity the model can close at the end of the repair trajectory.
DeepSeek achieves the {\em highest Final Fix Rate, reaching 99.31\%, indicating the strongest complete-repair capability}. Its GC is also the highest. While Gemini starts lower than DeepSeek, it also closes
the gap quite well (97.51\%), ending with the Final Fix Rate of 95.45\%.
%
Thus, {\em high GC reflects responsiveness to feedback and partial improvement, while high Final Fix reflects the ability to convert that progress into fully correct~solutions}.







As a sanity check, we compare controlled feedback with a {\em no-hint multi-turn baseline}. In the no-hint setting, models can revise their code for the same number of turns but receive no scenario-level hints. Gemini 2.5 Flash Lite reaches 78.63\% Final Fix Rate without hints, compared with 95.45\% under controlled feedback; Qwen3 Coder 30B reaches 57.72\%, compared with 90.90\%. These gains suggest that \tool's improvements are not merely due to repeated prompting, but are driven by repair-relevant hints. This is consistent with prior interactive-feedback benchmarks showing that feedback-guided repair improves over no-feedback interaction~\cite{pan2025benchmarkstalk,dai2025feedbackeval}.




\subsubsection{Do models use targeted hints effectively?}

Targeted Repair Success (TRS) measures whether a model fixes the failure scenario explicitly described by the current hint. 
DeepSeek achieves the highest TRS, indicating that it most reliably repairs the hinted failure regions. Gemini, GPT-4o-mini, and Qwen form the next group, suggesting that they can often act on targeted hints but do so less consistently than DeepSeek. In contrast, weaker models obtain lower TRS, showing that scenario-level feedback does not automatically translate into targeted repair. {\em These results demonstrate the value of separating targeted feedback use from final correctness}.

Hint Efficiency (HE) complements TRS by measuring how much assistance is needed to close a targeted scenario. A model with high TRS but low HE can repair hinted scenarios, but only after receiving more explicit guidance. Conversely, a model with high HE can close scenarios using shallower hints, indicating stronger diagnostic and refinement ability. DeepSeek ranks first in both TRS and HE, showing that it~not only repairs targeted scenarios most reliably, but also~often does so with less direct feedback. This combination distingui\-shes strong targeted repair from hint-dependent repair. GPT-4o-mini has slightly higher HE than Qwen, but Qwen has higher Final Fix, BP, and PMR. That contrast would~show that needing less assistance on closed scenarios does not necessarily imply better overall repair stability or final convergence.


\subsubsection{How much assistance do models need?}

\begin{figure}[t]
\centering
\includegraphics[width=0.95\linewidth]{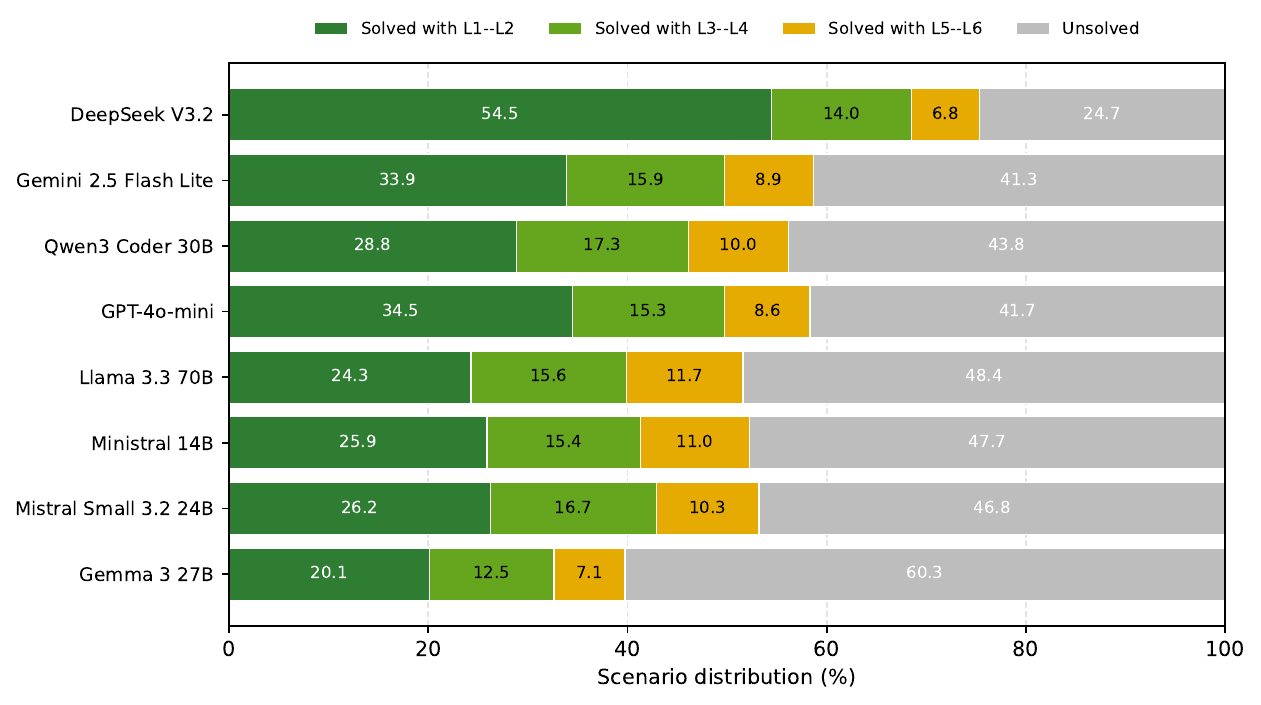}
\vspace{-12pt}
\caption{Hint-level success and fail distribution (RQ1).}
\label{fig:hint_level_distribution}
\end{figure}

Fig.~\ref{fig:hint_level_distribution} provides a finer-grained view of hint dependence by stratifying attempted scenarios according to the hint level at which they are closed. Percentages are computed over all attempted scenarios, including unresolved ones. DeepSeek closes the largest fraction of scenarios with shallow L1--L2 hints and leaves the smallest unresolved portion, showing that it can often infer repairs from weak diagnostic evidence. In contrast, weaker models leave larger unresolved portions or require deeper L5--L6 guidance, indicating greater dependence on explicit repair information. Thus, two models with similar final repair outcomes may still differ substantially in the amount of assistance they require.
This stratification supports our {\bf \em key idea of adaptive hint-depth control}: \tool does not merely record whether feedback helps, but identifies the level of assistance at which each model can make progress. Models that close more scenarios at L1--L2 demonstrate stronger diagnostic repair ability, whereas models that require L5--L6 hints or remain unresolved depend more on explicit guidance.

\subsubsection{Do models progressively close failure regions?}
We define a new metric, HintedClosedCoverage (HCC), to measure the fraction of initially failing test cases that were both exposed through progressive hints and closed by the final~solution. HCC assesses the effect of {failure-region control for progressive mechanism, i.e., a model also needs to close the exposed/hinted failure regions. Formally, let $F_0$ denote the set of tests failed by the initial solution, $H$ denote the set of failing tests selected for hints during the interaction, and $P_T$ denote the set of test cases passed by the final solution. We define
$
\mathrm{HintedClosedCoverage}
=
\frac{|F_0 \cap H \cap P_T|}{|F_0|}.
$
Fig.~\ref{fig:hinted-closed-coverage} shows that stronger models make more durable progress and close a larger fraction of the initially failing space through hinted regions. DeepSeek achieves the highest HCC at 20.45\%, followed by Qwen, GPT, etc. {\em The results show that {\bf \em failure-region control} provides a measurable way to track how much of the initial failure space is closed through controlled, scenario-level hints}.



\begin{figure}[t]
\centering
\includegraphics[width=0.9\linewidth]{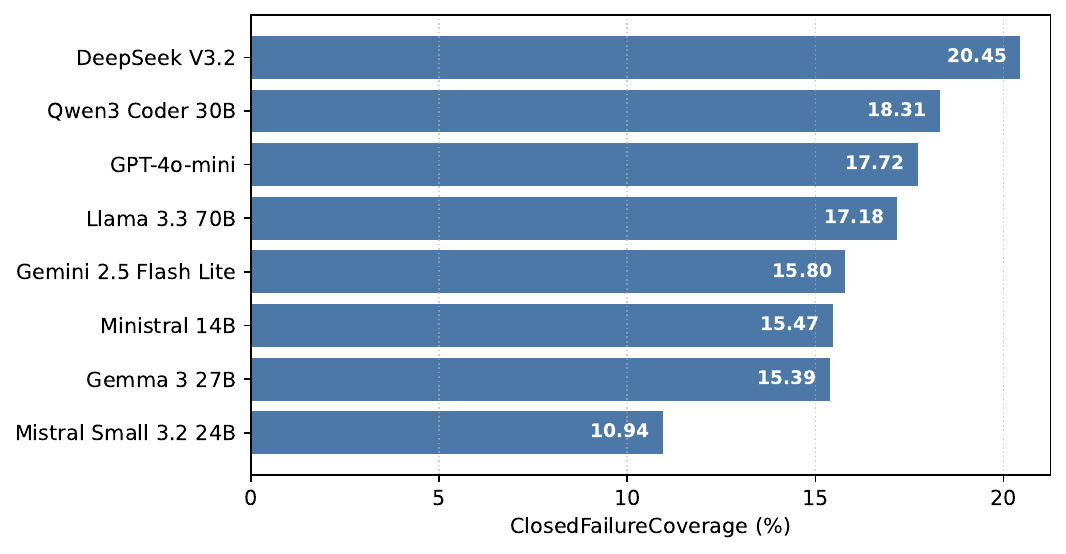}
\vspace{-12pt}
\caption{HintedClosedCoverage by candidate models (RQ1).}
\label{fig:hinted-closed-coverage}
\end{figure}



\subsubsection{Do targeted repairs generalize to other failures?}

Broader Repair Gain (BRG) measures whether a repair extends beyond the hinted scenario. BRG is much smaller than TRS for all models, suggesting that most improvements are localized to the targeted failure region. Gemini ranks first on BRG, followed by Ministral and DeepSeek, showing stronger ability to generalize from controlled hints to other hidden failures. The gap between TRS and BRG supports the need for progressive evaluation: {\em fixing one hinted scenario does not imply that the model has solved other failure regions, so the benchmark must expose and evaluate multiple regions over time}.



\begin{figure}[t]
\centering
\includegraphics[width=0.75\linewidth]{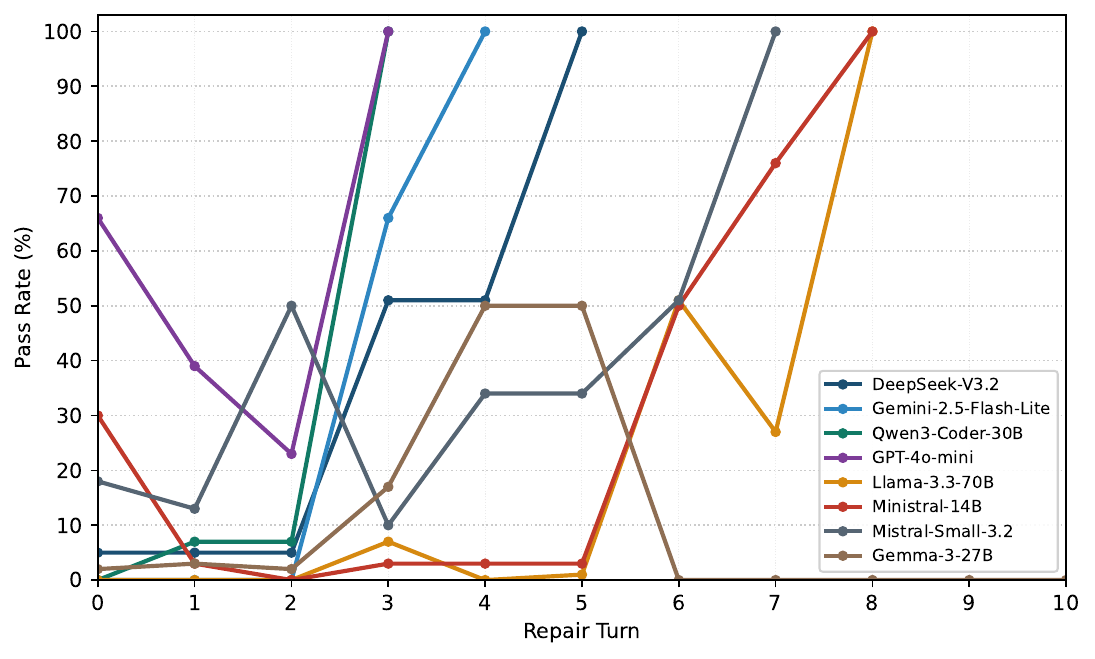} 
\vspace{-12pt}
\caption{Pass-rate trajectories for a single repair instance for candidate models.}
\label{fig:pass-rate-progress}
\end{figure}

\subsubsection{Are repair trajectories stable?}

Behavior Preservation (BP) and Progress Monotonicity Rate (PMR) measure whether models improve without destabilizing previously correct behavior. The two metrics are complementary. BP captures test-level regression by measuring if tests that previously passed remain passing after a repair, while PMR captures pass-rate-level stability by measuring if the overall pass rate decreases across turns. A model may maintain a non-decreasing pass rate while still changing which tests pass, so PMR alone can hide behavioral churn; BP helps expose such regressions.
DeepSeek and Qwen perform better on both BP and PMR, indicating that their repairs are not only effective but also stable across turns. 
In contrast, models with weaker preservation or monotonicity may still benefit from feedback, but their repair trajectories are less reliable since later edits can undo previously correct behavior. This shows that progressive repair should also measure
{\em whether the improvement is preserved during the~interaction}.

Fig.~\ref{fig:pass-rate-progress} illustrates how models can follow different repair trajectories on an instance. Stronger models reach full correctness in fewer turns: Qwen and GPT-4o-mini reach 100\% pass rate by turn 3. Weaker models require more turns and show less stable progress. Llama and Ministral eventually reach 100\%, but only after several low-progress or regressive turns. Gemma shows substantial partial progress but then regresses to 0\%.
This example highlights why {\tool} evaluates the full repair trajectory rather than only the final outcome.



\subsubsection{How efficiently do models converge?}


Because `Turns to Fix' is computed only over initially failing instances that are eventually repaired, it should be interpreted together with Final Fix Rate: a low Turns to Fix indicates fast convergence only among solved instances, not necessarily stronger repair coverage overall. DeepSeek ranks first on Turns to Fix while also achieving the highest Final Fix Rate, indicating that it can translate feedback into complete repairs both reliably and quickly. Gemini and GPT-4o-mini also require relatively few turns, suggesting efficient use of feedback on instances they successfully repair. In contrast, Qwen reaches a high Final Fix Rate but requires more turns, indicating strong~final repair ability but weaker convergence efficiency. This distinction matters for {\em progressive repair workflows}, where repeated feedback turns impose cost even when the final repair succeeds.

\begin{figure}[t]
\centering
\includegraphics[width=0.75\linewidth]{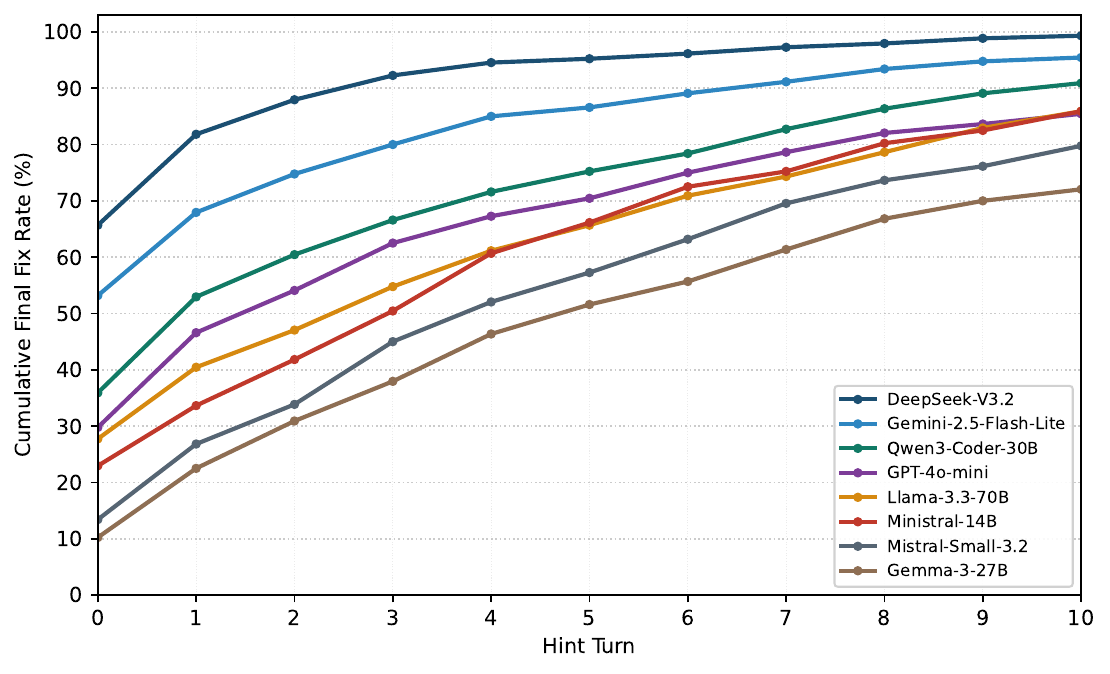}
\vspace{-12pt}
\caption{Progressive improvement in cumulative Final Fix Rate (RQ1).}
\label{fig:cumulative-fix-rate}
\end{figure}


Fig.~\ref{fig:cumulative-fix-rate} reports the cumulative Final Fix Rate by turn (over 10 feedback turns), showing both convergence speed and final repair coverage: a steeper early rise shows that a model quickly converts feedback into complete repairs. DeepSeek starts from the highest Initial Fix Rate, rises sharply in the first few turns, and approaches near-complete repair by the end of the budget. Gemini follows a similar but lower trajectory, with steady early gains and strong final coverage. The other models improve more gradually. Qwen starts much lower than Gemini but steadily catches up to the third-highest final coverage, indicating strong eventual repair ability. 
Mistral Small and Gemma remain lower throughout, suggesting that feedback helps them improve but less often leads to complete repair.
%



\subsubsection{Which models are strongest overall?}

Table~\ref{tab:rq1_metric_ranking} summarizes the metric-wise ranking. We do not treat the average rank as a principled aggregate score.
Instead, we use it to provide a descriptive view of {\bf \em model repair profile}.
DeepSeek is the strongest all-around model, ranking first in most metrics and combining high initial correctness, effective targeted repair, strong preservation, high hint efficiency, high final repair, and fast convergence. Gemini and Qwen form the next tier with different strengths: Gemini is stronger in broader repair and net improvement, while Qwen is stronger in preservation, monotonicity, and final repair. Other models show substantial Gap Closure but lower Final Fix Rates, weaker stability, or longer convergence, suggesting that they benefit from feedback but less consistently convert partial progress into reliable~full repair. Overall, this shows that good interactive repair depends on final correctness, and also on targeted repair, progressive improvement, stability, hint efficiency, and convergence.

\subsection{Robustness of Controlled Feedback (RQ2)}
\label{sec:rq2}


\subsubsection{Setting and Metrics}

RQ2 evaluates whether controlling feedback generation makes interactive program-repair evaluation more stable. We compare two settings that use the same initial candidate programs, evaluated models, turn budget, execution environment, and feedback-generation model; the only difference is the feedback protocol. \emph{Controlled Feedback} is {\tool}.
\emph{Vanilla Feedback} is an ablated baseline inspired by prior interactive feedback evaluation~\cite{pan2025benchmarkstalk}, where the feedback model generates a hint at each turn without scenario-level testcase selection or hint-depth control.




Both settings start from the same initial candidate repairs, so the Initial Fix Rate is identical for each model. The~comparison thus focuses on the interactive stage after feedback is introduced. For each setting, we run the benchmark three times and report the mean and standard deviation. Since Vanilla Feedback does not define targeted scenarios or hint levels, we use metrics shared by both settings: Repair Rate, Preservation Rate, Final Fix Rate, and Turns to Fix. We define the turn-level Repair Rate as the fraction of previously failing tests that become passing after the current repair
$ \mathrm{RepairRate}_{i,t} =
\frac{|F_{i,t-1}\cap P_{i,t}|}{|\mathcal{T}_i|}.
$
Final Fix Rate measures the fraction of instances that are fully repaired within the  budget, and Turns to Fix measures convergence speed among fully repaired instances. 



\begin{table}[t]
\centering
\scriptsize
\setlength{\tabcolsep}{1.5pt}
\caption{Robustness comparison between Controlled Feedback (Ours) and Vanilla Feedback across three repeated runs (RQ2).}
\label{tab:rq2_robustness}
\vspace{-6pt}
\resizebox{0.49\textwidth}{!}{%
\vspace{-3pt}
\begin{tabular}{l c l c c c c}
\toprule
\textbf{Model} &
\textbf{Init.Fix$\uparrow$} &
\textbf{Setting} &
\textbf{Repair Rate$\uparrow$} &
\textbf{Preserve.Rate$\uparrow$} &
\textbf{Final Fix$\uparrow$} &
\textbf{Turns to Fix$\downarrow$} \\
\midrule

\multirow{2}{*}{DeepSeek}
& \multirow{2}{*}{65.68}
& Control
& 23.91 $\pm$ \textbf{0.18}
& 98.27 $\pm$ \textbf{0.06}
& 99.18 $\pm$ \textbf{0.16}
& 2.57 $\pm$ \textbf{0.05}  \\
&
& Vanilla
& 32.70 $\pm$ 1.24
& 98.75 $\pm$ 0.31
& 96.44 $\pm$ 0.82
& 2.05 $\pm$ 0.14  \\

\midrule

\multirow{2}{*}{Gemini}
& \multirow{2}{*}{53.18}
& Control
&  27.74 $\pm$ \textbf{0.31}
&  93.72 $\pm$ \textbf{0.12}
&  95.31 $\pm$ \textbf{0.21}
&  3.24 $\pm$ \textbf{0.07}  \\
&
& Vanilla
& 40.90 $\pm$ 1.75
& 95.08 $\pm$ 0.52
& 97.50 $\pm$ 0.65
& 2.23 $\pm$ 0.16 \\

\midrule

\multirow{2}{*}{Qwen}
& \multirow{2}{*}{35.90}
& Control
&  23.06 $\pm$ \textbf{0.20}
&  95.14 $\pm$ \textbf{0.10}
&  90.76 $\pm$ \textbf{0.19}
&  3.84 $\pm$ \textbf{0.06}  \\
&
& Vanilla
& 33.76 $\pm$ 0.61
& 95.73 $\pm$ 0.58
& 93.94 $\pm$ 0.76
& 2.58 $\pm$ 0.13 \\

\midrule

\multirow{2}{*}{GPT}
& \multirow{2}{*}{29.77}
& Control
&  18.28 $\pm$ \textbf{0.17}
&  94.86 $\pm$ \textbf{0.09}
&  85.32 $\pm$ \textbf{0.18}
&  3.70 $\pm$ \textbf{0.05}  \\
&
& Vanilla
& 28.85 $\pm$ 0.72
& 95.69 $\pm$ 0.61
& 90.30 $\pm$ 0.51
& 2.41 $\pm$ 0.12 \\

\midrule

\multirow{2}{*}{Llama}
& \multirow{2}{*}{27.72}
& Control
&  20.07 $\pm$ \textbf{0.28}
&  93.68 $\pm$ \textbf{0.14}
&  85.64 $\pm$ \textbf{0.31}
&  4.36 $\pm$ \textbf{0.08}  \\
&
& Vanilla
& 30.28 $\pm$ 1.42
& 92.06 $\pm$ 0.70
& 90.84 $\pm$ 4.28
& 2.92 $\pm$ 0.25 \\

\midrule

\multirow{2}{*}{Ministral}
& \multirow{2}{*}{22.95}
& Control
&  21.39 $\pm$ \textbf{0.23}
&  91.31 $\pm$ \textbf{0.13}
&  85.72 $\pm$ \textbf{0.25}
&  4.29 $\pm$ \textbf{0.07}  \\
&
& Vanilla
& 32.67 $\pm$ 0.94
& 92.20 $\pm$ 0.43
& 91.21 $\pm$ 0.82
& 2.78 $\pm$ 0.12 \\

\midrule

\multirow{2}{*}{Mistral}
& \multirow{2}{*}{13.40}
& Control
&  20.66 $\pm$ \textbf{0.25}
&  92.68 $\pm$ \textbf{0.11}
&  79.94 $\pm$ \textbf{0.28}
&  4.35 $\pm$ \textbf{0.06}  \\
&
& Vanilla
& 31.18 $\pm$ 0.74
& 95.31 $\pm$ 0.34
& 85.30 $\pm$ 1.52
& 2.85 $\pm$ 0.13 \\

\midrule

\multirow{2}{*}{Gemma}
& \multirow{2}{*}{10.22}
& Control
&  17.58 $\pm$ \textbf{0.27}
&  90.26 $\pm$ \textbf{0.15}
&  72.18 $\pm$ \textbf{0.32}
&  4.29 $\pm$ \textbf{0.08}  \\
&
& Vanilla
& 27.42 $\pm$ 1.36
& 91.47 $\pm$ 0.78
& 80.18 $\pm$ 2.35
& 2.96 $\pm$ 0.21  \\

\bottomrule
\end{tabular}%
}
\end{table}

\subsubsection{Robustness Comparison}

Table~\ref{tab:rq2_robustness} reports a repeated-run stability analysis over three runs. While this number of runs limits strong statistical claims about variance, the pattern is consistent: Controlled Feedback yields lower run-to-run variation than Vanilla Feedback in every model-metric comparison across all eight models and four shared metrics. Averaged across models, the standard deviation of Repair Rate decreases from 1.10 to 0.24, Preservation Rate from 0.53 to 0.11, Final Fix Rate from 1.46 to 0.24, and Turns to Fix from 0.16 to 0.07 when moving from Vanilla to Controlled Feedback. These correspond to relative reductions of approximately 78\%, 79\%, 84\%, and 56\%, respectively. Thus, the repeated-run results support that controlling the targeted failure region and hint depth improves the stability of interactive repair evaluation.





\subsubsection{Statistical Testing}

We measure ranking stability with Kendall's $\tau$ by ranking models independently in each repeated run and averaging $\tau$ over all run pairs; for Turns to Fix, lower values are ranked higher. From Table~\ref{tab:rq2-kendall}, Controlled Feedback produces more stable rankings than Vanilla Feedback across all shared metrics, with a higher average $\tau$ ($0.982$ vs. $0.905$). This indicates that {\tool} can reduce metric variance as well as preserve model ordering across repeated evaluations.




\subsubsection{Mean performance versus stability}

Vanilla Feedback often gives higher mean Repair Rate, higher mean Final Fix Rate, and fewer Turns to Fix. We do not interpret this as evidence that Vanilla Feedback is a better evaluation. Because Vanilla Feedback does not control the targeted failure region or hint depth, different runs may expose different failure modes or reveal different levels of repair directness. This can make some runs easier, but also makes the measured outcome more dependent on feedback randomness. In contrast, Controlled Feedback has lower variance and clearer attribution: each repair attempt is tied to a known failure scenario and a known hint-depth level. Thus, RQ2 supports Controlled Feedback as a more stable protocol for measuring interactive APR capability. 




\begin{table}[t]
\centering
\caption{Ranking Stability by Kendall's $\tau$ Statistical Test.}
\label{tab:rq2-kendall}
\vspace{-6pt}
\resizebox{0.7\columnwidth}{!}{
\begin{tabular}{lccccc}
\toprule
\textbf{Setting} & \textbf{Repair} & \textbf{Preserve} & \textbf{Final} & \textbf{Turns} & \textbf{Avg.} \\
\midrule
Controlled & \textbf{1.000} & \textbf{1.000} & \textbf{1.000} & \textbf{0.928} & \textbf{0.982} \\
Vanilla    & 0.952 & 0.952 & 0.857 & 0.857 & 0.905 \\
\bottomrule
\end{tabular}
}
\end{table}
\subsection{Hyperparameter Sensitivity (RQ3)}
\label{sec:rq3}

\newcommand{\rankone}[1]{\cellcolor{green!25}\textbf{#1}}
\newcommand{\ranktwo}[1]{\cellcolor{blue!18}\textbf{#1}}
\newcommand{\rankthree}[1]{\cellcolor{orange!22}\textbf{#1}}


\begin{table}[t]
\centering
\scriptsize
\setlength{\tabcolsep}{4pt}
\renewcommand{\arraystretch}{1.18}
\caption{Hyperparameter sensitivity results (RQ3). \textcolor{green!50!black}{Green}, \textcolor{blue!70!black}{blue}, and \textcolor{orange!80!black}{orange} mark the top-1, top-2, and top-3 values.}
\label{tab:sensitivity}
\vspace{-6pt}
\resizebox{0.5\textwidth}{!}{
\begin{tabular}{p{2.6cm} l c c c c}
\toprule
\multirow{1}{*}{\textbf{Setting}} 
& \multirow{1}{*}{\textbf{Model}} 
& \multirow{1}{*}{\textbf{Initial Fix$\uparrow$}}
& \multirow{1}{*}{\textbf{Final Fix$\uparrow$}} 
& \textbf{Prog. Mono.$\uparrow$} 
& \textbf{Hint Eff.$\uparrow$} \\
\midrule

\multirow{3}{=}{Main setting: $\alpha=10,\beta=3$}
& Gemini & \rankone{53.18} & \rankone{95.45} & \rankone{81.39} & \rankone{3.76} \\
& Llama & \ranktwo{27.72} & \ranktwo{85.90} & \rankthree{71.61} & \ranktwo{3.21} \\
& Mistral & \rankthree{13.40} & \rankthree{79.77} & \ranktwo{76.35} & \rankthree{3.02} \\
\midrule

\multirow{3}{=}{Larger budget: $\alpha=15,\beta=4$}
& Gemini & \rankone{53.18} & \rankone{98.18} & \rankone{83.81} & \rankone{3.64} \\
& Llama & \ranktwo{27.72} & \ranktwo{94.77} & \rankthree{73.38} & \ranktwo{3.07} \\
& Mistral & \rankthree{13.40} & \rankthree{90.22} & \ranktwo{79.00} & \rankthree{2.95} \\
\midrule

\multirow{3}{=}{Smaller scenario budget: $\alpha=10,\beta=2$}
& Gemini & \rankone{53.18} & \rankone{96.13} & \rankone{79.90} & \rankone{3.45} \\
& Llama & \ranktwo{27.72} & \ranktwo{79.09} & \rankthree{71.68} & \ranktwo{2.90} \\
& Mistral & \rankthree{13.40} & \rankthree{71.81} & \ranktwo{78.71} & \rankthree{2.65} \\
\midrule

\multirow{3}{=}{Alternative feedback \\ model: DeepSeek V3.2}
& Gemini & \rankone{53.18} & \rankone{92.04} & \rankone{82.54} & \rankone{3.28} \\
& Llama & \ranktwo{27.72} & \ranktwo{75.00} & \rankthree{73.29} & \ranktwo{2.93} \\
& Mistral & \rankthree{13.40} & \rankthree{67.50} & \ranktwo{77.81} & \rankthree{2.53} \\

\bottomrule
\end{tabular}
}
\end{table}

RQ3 evaluates whether \tool's model comparisons are robust to varied hyperparameters. We focus on three representative models from different performance tiers: Gemini, Llama, and Mistral. We vary the total interaction budget~($\alpha$), the per-scenario refinement budget ($\beta$), and the feedback model (DeepSeek V3.2). As seen in Table~\ref{tab:sensitivity}, the relative~or-dering is largely stable across settings: Gemini consistently ranks highest in Final Fix Rate and Hint Efficiency, followed by Llama and Mistral, yielding Kendall's $\tau$=1.0. This suggests that the ranking is not an artifact of a single budget choice or feedback generator. Increasing the budget improves Final Fix Rate for all models, especially Llama and Mistral, while reducing the per-scenario budget mainly hurts these weaker models and lowers Hint Efficiency. This suggests that additional refinement opportunities help weaker models, whereas limiting scenario-level refinement makes targeted repair harder.

Using DeepSeek V3.2 as the feedback model preserves the same relative ordering, but lowers Final Fix Rate and~Hint Efficiency compared with the main setting, especially for Llama and Mistral. Interestingly, Progress Monotonicity remains comparable or slightly higher. Overall, these results show that rankings are reasonably stable under the tested~parameter changes, while absolute performance reflects the interaction budget and the amount and quality of the feedback~source.

\vspace{-2pt}
\section{Concluding Remarks}
\tool reveals that model improvement is multidimensional. Models differ not only in whether they eventually pass all tests, but also in how precisely they use targeted hints, how much assistance they require, whether repairs generalize beyond the hinted scenario, and whether progress remains stable across turns. Our future work includes adding the adaptive process to the question level
with easy, medium, and hard problems, and adaptively challenging the target model.


\balance

\bibliographystyle{IEEEtran}

\bibliography{references,apr-references}

\end{document}